\documentclass[a4paper, conference]{IEEEtran}

\usepackage{cite}
\usepackage{amsmath,amssymb,amsfonts}
\usepackage{algorithmic}
\usepackage{graphicx}
\usepackage{textcomp}
\usepackage{xcolor}
\usepackage{paralist}
\usepackage{eurosym}
\usepackage{flushend}
\usepackage{url}
\usepackage[utf8]{inputenc}
\usepackage[colorlinks=true]{hyperref}
\usepackage{soul}
\usepackage{caption}
\usepackage[list=true]{subcaption}
\usepackage{booktabs}
\usepackage{tabu}
\usepackage{wrapfig}
\usepackage{arydshln}
\usepackage[normalem]{ulem}
\usepackage{multirow}
\usepackage{siunitx}
\usepackage{adjustbox}
\usepackage{array}
\usepackage{diagbox}
\usepackage{xfrac}

\newcommand\blfootnote[1]{%
  \begingroup
  \renewcommand\thefootnote{}\footnote{#1}%
  \addtocounter{footnote}{-1}%
  \endgroup
}

\begin{document}

\title{Receiver-Aware Analysis and Verification of the Spectral Separation Coefficient Under Interference-Induced Degradation}

\author{\IEEEauthorblockN{Lucas Heublein, Fabian Benschuh, Alexander Rügamer, Felix Ott}
  \IEEEauthorblockA{Fraunhofer Institute for Integrated Circuits IIS, 90411 Nürnberg, Germany}
  \IEEEauthorblockA{\{lucas.heublein, fabian.benschuh, alexander.ruegamer, felix.ott\}@iis.fraunhofer.de}
}

\maketitle

\begin{abstract}
Interference poses a significant challenge to satellite-based positioning systems, making it essential to accurately quantify the effects of specific interference types on receiver performance and the resulting reliability of position computation. In current practice, interference effects are often quantified using receiver-independent metrics, with receiver-specific front-end characteristics either idealized or only implicitly considered. In this paper, we address this limitation by explicitly incorporating receiver-specific front-end characteristics into the computation of interference effects and validating the resulting receiver-dependent analysis experimentally. Therefore, we record a real-world open-field dataset comprising 210 distinct interference scenarios and compute the receiver-dependent spectral separation coefficient (SSC) and interference impact for a specific receiver module. Furthermore, we verify the computation using a controlled dataset generated with a radio frequency constellation simulator (RFCS), employing the same receiver module and replaying similar interferences classes. The comparison of results obtained in both environments demonstrates the robustness of the interference impact computation.
\end{abstract}
\IEEEpeerreviewmaketitle

\section{Introduction}
\label{label_introduction}

Jamming devices disrupt signals from GNSS and pose a significant threat by compromising the robustness and reliability of accurate positioning~\cite{ferre_fuente,swinney_woods}. This challenge has intensified in recent years due to the growing availability of easily accessible jamming devices~\cite{mehr_minetto_dovis,miguel_chen_lo}. The detection~\cite{brieger_ion_gnss}, classification~\cite{ott_heublein_icl}, characterization~\cite{heublein_feigl_crpa}, localization~\cite{raichur_ion_gnss,heublein_wielenberg}, and mitigation~\cite{chen_liu_huang} of interference have become critical research topics. Accurate classification of interference waveforms, which serve as distinctive fingerprints of jamming devices, enables the identification of jammer intent~\cite{mehr_dovis_TAES}. A wide range of techniques have been proposed to address these challenges, including classical signal processing methods~\cite{yang_kang,murrian_narula,gross_humphreys} as well as machine learning-based approaches~\cite{raichur_heublein,heublein_feigl_jispin,manjunath_heublein}.

Accurate navigation in the presence of interference requires a precise characterization of how ongoing interference affects receiver performance. Quantifying the influence of different interference types on navigation reliability is therefore a prerequisite for robust positioning under adverse signal conditions~\cite{kaplan,betz_ion,ruegamer_dissertation}. The influence of interference on the receiver's carrier-to-noise density ratio ($C/N_0$) can be computed directly from raw baseband samples by estimating the interference power and its spectral characteristics~\cite{kaplan}. Alternatively, time–frequency representations such as spectrograms can be used to characterize the temporal and spectral structure of interference and to quantify its overlap with the desired signal spectrum~\cite{heublein_feigl_crpa}. The interference impact is obtained by evaluating spectral overlap measures or equivalent noise contributions that relate interference characteristics to receiver performance metrics such as signal-to-noise-and-interference-ratio (SNIR) or effective $C\!/\!N_0$~\cite{betz_ion,betz_goldstein,wallner_hein,spilker_axelrad}. These approaches enable the assessment of interference effects without requiring explicit knowledge of the interference source, relying instead on observable signal properties~\cite{ahmed_mohamadi}.

The SSC constitutes a compact and physically motivated metric for quantifying interference impact, as it explicitly captures the effective spectral overlap between interference and the desired signal at the receiver correlator and thus parameterizes the dominant mechanism governing interference-induced degradation of receiver performance~\cite{kaplan,betz_goldstein}. However, existing formulations predominantly employ receiver-independent or idealized receiver models, in which front-end filtering, bandwidth limitations, and implementation-specific processing effects are either abstracted or implicitly assumed~\cite{wallner_hein}. In this work, we use a receiver-aware SSC and impact formulation that explicitly incorporates receiver front-end characteristics and validate the resulting interference impact assessment using both real-world measurements and controlled simulations on a fixed receiver architecture.

\textbf{Contributions.} The primary objective of this work is to compute the SSC as a function of receiver front-end parameters and to analyze and validate this computation using both real-world measurements and simulated data. The contributions of this paper are as follows: (1) We review the theoretical preliminaries of correlator SNIR, the effective $C\!/\!N_0$, and the jamming resistance formulation. (2) Therefore, we use the SSC to explicitly account for receiver front-end filtering. (3) Using a representative receiver module, we demonstrate the receiver-dependent computation of the SSC. (4) We present a real-world open-field measurement dataset, and (5) experimentally demonstrate the applicability of the receiver-aware SSC. (6) Finally, we validate the SSC computation through controlled experiments using signals generated with an RFCS and processed by the same receiver.

\section{Related Work}
\label{label_related_work}

The theoretical foundations of GNSS signal reception were established in early works such as Spilker et al.~\cite{spilker_axelrad}, which developed correlator-based receiver models and spectral representations that underpin modern interference analysis. Betz~\cite{betz,betz_ion} analyzed the impact of narrowband and partial-band interference on receiver performance and $C\!/\!N_0$ estimation, establishing spectral overlap integrals as the key mechanism later encapsulated by the SSC. Several studies have employed SSC or closely related spectral overlap measures primarily for signal design and compatibility analysis. Building on this framework, Kaplan and Hegarty~\cite{kaplan} derived the correlator output SNIR and formalized the SSC as a spectral overlap metric governing effective $C\!/\!N_0$ degradation under interference. Betz \& Goldstein~\cite{betz_goldstein} introduced the jamming resistance quality factor to compare candidate GNSS signal designs, while Issler et al.~\cite{issler_ries} provided empirical spectral measurements demonstrating that wider transmitted bandwidths reduce mutual interference. Modulation-specific analyses, such as Pratt \& Owen~\cite{pratt_owen}, investigated how binary offset carrier (BOC) waveform properties influence spectral overlap, and Wallner et al.~\cite{wallner_hein} applied overlap-based metrics to quantify mutual interference between GPS and Galileo signals. In these works, SSC-related quantities are primarily used as design-time or compatibility metrics rather than as receiver-specific performance predictors.

More recent contributions have extended SSC-based analysis toward practical interference evaluation. Rügamer~\cite{ruegamer_dissertation} clarified the receiver-dependent nature of SSC by explicitly incorporating front-end filtering and bandwidth effects, linking SSC to effective $C\!/\!N_0$ degradation in real receivers. Subsequent studies have used SSC directly to compute interference impact under jamming scenarios, including directional jammer analysis~\cite{bacic_sugar}, receiver-filter-weighted effective $C\!/\!N_0$ modeling~\cite{ahmed_mohamadi}, and compatibility assessments between GNSS and emerging LEO-based navigation signals~\cite{yan_wang_bian}. SSC has also been employed as a comparative metric in waveform and signal design studies~\cite{deng_yang_qian,lee_han_won}. Despite its widespread use, the limitations of SSC have also been examined. Enneking et al.~\cite{enneking_antreich} showed that while SSC captures interference-induced noise effects, it does not fully characterize discriminator distortion and tracking bias in certain CDMA (code division multiple access) scenarios, highlighting the importance of receiver-specific context when interpreting SSC values.

In contrast to existing literature, which primarily applies SSC for signal compatibility analysis, waveform comparison, or generalized interference assessment, this work focuses on the influence of interference on a specific receiver architecture. We use the SSC as a receiver-aware explanatory variable to analyze measured interference effects, emphasizing its role in mapping spectral interference characteristics to observable receiver performance degradation rather than as a purely abstract or design-level metric. This receiver-centric perspective enables a direct assessment of interference impact under practical receiver implementations.
\section{Spectral Separation Coefficient}
\label{label_preliminaries}

The performance of signal acquisition, carrier tracking, and data demodulation depends critically on the SNIR at the output of the receiver correlator. Consequently, the effect of radio-frequency (RF) interference can be quantified by evaluating its impact on the correlator output SNIR. Section~\ref{label_preliminaries_SNIR} introduces the correlator SNIR and the effective carrier-to-noise density ratio. Section~\ref{label_preliminaries_jamming_resistance} presents the jamming resistance formulation and the SSC. Section~\ref{label_preliminaries_receiver} extends the SSC to account for receiver filtering and relates it to observable receiver performance. Table~\ref{table_notations} gives an overview of all notations.

\renewcommand{\arraystretch}{0.93}
\begin{table}[t!]
\begin{center}
\setlength{\tabcolsep}{3.5pt}
    \caption{Overview of notations used in the SSC analysis.}
    \label{table_notations}
    \vspace{-0.15cm}
    \begin{tabular}{ p{0.5cm} | p{0.5cm} }
    \multicolumn{1}{c|}{\textbf{Notation}} & \multicolumn{1}{c}{\textbf{Description}} \\ \hline
    \multicolumn{1}{c|}{$\rho_c$} & \multicolumn{1}{l}{Signal-to-noise-plus-interference ratio (SNIR)} \\
    \multicolumn{1}{c|}{$T$} & \multicolumn{1}{l}{Coherent integration time of the correlator [$s$]} \\
    \multicolumn{1}{c|}{$f$} & \multicolumn{1}{l}{Frequency variable [Hz]} \\
    \multicolumn{1}{c|}{$B_r$} & \multicolumn{1}{l}{Effective receiver bandwidth [Hz]} \\
    \multicolumn{1}{c|}{$H_R(f)$} & \multicolumn{1}{l}{Receiver transfer function (dimensionless)} \\
    \multicolumn{1}{c|}{$C_s$} & \multicolumn{1}{l}{Received power of the signal over bandwidth [W]} \\
    \multicolumn{1}{c|}{$C_i$} & \multicolumn{1}{l}{Received power of the interference signal over BW [W]} \\
    \multicolumn{1}{c|}{$N_0$} & \multicolumn{1}{l}{One-sided PSD of additive white Gaussian noise [$\text{W}\!/\text{Hz}$]} \\
    \multicolumn{1}{c|}{$S_s(f)$} & \multicolumn{1}{l}{PSD of the desired signal [$\text{1}\!/\text{Hz}$]} \\
    \multicolumn{1}{c|}{$S_i(f)$} & \multicolumn{1}{l}{PSD of the interference signal [$\text{1}\!/\text{Hz}$]} \\
    \multicolumn{1}{c|}{$\tau$} & \multicolumn{1}{l}{Code delay between local replica and received signal [$s$]} \\
    \multicolumn{1}{c|}{$C_s/\!N_0$} & \multicolumn{1}{l}{Carrier-to-noise density ratio of the desired signal [Hz]} \\
    \multicolumn{1}{c|}{$(C_s/\!N_0)_{\mathrm{eff}}$} & \multicolumn{1}{l}{Effective $C\!/\!N_0$ density ratio during interference [Hz]} \\
    \multicolumn{1}{c|}{$R_c$} & \multicolumn{1}{l}{Spreading code rate [chips/$s$]} \\
    \multicolumn{1}{c|}{$Q$} & \multicolumn{1}{l}{Jamming resistance quality factor (dimensionless)} \\
    \multicolumn{1}{c|}{$\kappa_{is}$} & \multicolumn{1}{l}{SSC between interference and desired signal [$s$ or $\text{1}\!/\text{Hz}$]} \\
    \end{tabular}
\end{center}
\end{table}

\subsection{Correlator SNIR and Effective $C_s/\!N_0$}
\label{label_preliminaries_SNIR}

When the aggregate interference can be modeled as statistically stationary, and when the spectra of the desired signal and/or interference can be approximated as locally linear over the reciprocal of the correlator integration time, the prompt correlator output SNIR is given by~\cite{betz}
\begin{equation}
\label{eq:correlator-snir}
\resizebox{0.99\linewidth}{!}{$
    \rho_c
    =
    \frac{
    2 T \frac{C_s}{N_0}
    \left[
    \displaystyle
    \int_{-B_r/2}^{B_r/2}
    S_s(f) e^{j 2 \pi f \tau} \, df
    \right]^2
    }{
    \displaystyle
    \int_{-B_r/2}^{B_r/2}
    \left| H_R(f) \right|^2 S_s(f) \, df
    \;+\;
    \frac{C_i}{N_0}
    \int_{-B_r/2}^{B_r/2}
    \left| H_R(f) \right|^2 S_i(f) S_s(f) \, df
    },
$}
\end{equation}
where $T$ denotes the coherent integration time of the correlator in $s$, $C_s$ is the received power of the desired signal (in watts) over infinite bandwidth (BW), and $N_0$ is the one-sided power spectral density of additive white Gaussian noise (in $\text{W}\!/\text{Hz}$)~\cite{kaplan}. The function $H_R(f)$ represents the receiver transfer function (ratio), while $S_s(f)$ is the normalized power spectral density (PSD) of the transmitted signal (in $\text{W}\!/\text{Hz}$), satisfying
\begin{equation}
    \int_{-B_r/2}^{B_r/2} S_s(f) \, df = 1.
\end{equation}
The quantity $C_i$ denotes the received interference power (in watts), and $S_i(f)$ is the normalized PSD (in $\text{W}\!/\text{Hz}$) of the aggregate interference over infinite bandwidth~\cite{kaplan}.

Eq.~\ref{eq:correlator-snir} shows that RF interference degrades correlator performance through a spectral interaction term involving the product $S_i(f) S_s(f)$ weighted by the receiver filter response. This interaction motivates the definition of spectral separation measures that quantify the degree of overlap between the desired signal spectrum and the interference spectrum.

The quality of a received GNSS signal is traditionally characterized by its carrier-to-noise density ratio $C\!/\!N_0$, under the assumption that the disturbance is well modeled as additive white noise and can therefore be represented by a scalar noise density. However, as Eq.~\ref{eq:correlator-snir} demonstrates, the presence of non-white interference necessitates an explicit spectral description, since both its total power and its spectral shape influence the correlator output SNIR. Direct analysis of the combined noise and interference contributions in the spectral domain is often analytically cumbersome. A common and effective alternative is to introduce an equivalent fictitious white noise process whose power spectral density yields the same correlator output SNIR as the actual combination of thermal noise and interference. This construction allows the degradation induced by interference to be captured as an effective increase in noise density, enabling straightforward performance analysis while preserving the impact of spectral overlap between the desired signal and the interference~\cite{kaplan}.

To derive an effective carrier-to-noise density ratio, denoted $(C_s/\!N_0)_{\mathrm{eff}}$, we first consider the interference-free case. In the absence of interference and assuming infinite receiver bandwidth, Eq.~\ref{eq:correlator-snir} reduces to the signal-to-noise ratio (SNR) at the prompt correlator output $\rho_c = 2 T \frac{C_s}{N_0}$. Equivalently, the carrier-to-noise density ratio can be expressed in terms of the correlator output SNR as $\frac{C_s}{N_0} = \frac{\rho_c}{2T}$. When both thermal noise and interference are present, the effective carrier-to-noise density ratio is defined analogously, but using the correlator output SNIR given by Eq.~\ref{eq:correlator-snir}~\cite{kaplan}. Specifically,
\begin{equation}
\label{eq:cn0-eff-def}
    \left( \frac{C_s}{N_0} \right)_{\mathrm{eff}}
    =
    \frac{\rho_c}{2T}.
\end{equation}
Substituting Eq.~\ref{eq:correlator-snir} into Eq.~\ref{eq:cn0-eff-def} yields
\begin{equation}
\label{eq:cn0-eff-expanded}
\resizebox{0.99\linewidth}{!}{$
    \left( \frac{C_s}{N_0} \right)_{\mathrm{eff}}
    =
    \left( \frac{C_s}{N_0} \right)
    \frac{
    \left[ \displaystyle \int_{-B_r/2}^{B_r/2} S_s(f)\, df \right]^2
    }{
    \displaystyle
    \int_{-B_r/2}^{B_r/2} S_s(f)\, df
    \;+\;
    \frac{C_i}{N_0}
    \int_{-B_r/2}^{B_r/2} S_i(f) S_s(f)\, df
    }.
$}
\end{equation}
Rearranging Eq.~\ref{eq:cn0-eff-expanded} gives~\cite{kaplan}
\begin{equation}
\label{eq:cn0-eff-final}
    \left( \frac{C_s}{N_0} \right)_{\mathrm{eff}}
    =
    \frac{
    \displaystyle \int_{-B_r/2}^{B_r/2} S_s(f)\, df
    }{
    \displaystyle
    \frac{1}{(C_s/N_0)}
    \;+\;
    \frac{C_i}{C_s}
    \int_{-B_r/2}^{B_r/2} S_i(f) S_s(f)\, df
    }.
\end{equation}

\subsection{Jamming Resistance and SSC}
\label{label_preliminaries_jamming_resistance}

According to Betz et al.~\cite{betz_goldstein}, Eq.~\ref{eq:cn0-eff-final} can be equivalently expressed in a more compact form as
\begin{equation}
\label{eq:cn0-eff-Q}
    \left( \frac{C_s}{N_0} \right)_{\mathrm{eff}}
    =
    \frac{
    \displaystyle \int_{-B_r/2}^{B_r/2} S_s(f)\, df
    }{
    \displaystyle
    \frac{1}{(C_s/N_0)}
    \;+\;
    \frac{C_i}{C_s}\,\frac{1}{Q R_c}
    },
\end{equation}
where $C_s/\!N_0$ denotes the unjammed carrier-to-noise density ratio of the received signal prior to receiver filtering, and $C_i/C_s$ is the jammer-to-signal power ratio at the receiver input. The quantity $Q$ is a dimensionless jamming resistance quality factor that depends on the spectral characteristics of both the interference and the desired signal, while $R_c$ is the spreading code rate in chips per second~\cite{kaplan}.

Increasing the value of $Q$ improves the effective carrier-to-noise density ratio, $(C_s/\!N_0)_{\mathrm{eff}}$, indicating enhanced resistance to interference. Consequently, higher values of $Q$ correspond to reduced jammer effectiveness. By comparing Eq.~\ref{eq:cn0-eff-final} and Eq.~\ref{eq:cn0-eff-Q}, the jamming resistance quality factor is identified as
\begin{equation}
\label{eq:Q-def}
    Q
    =
    \frac{
    \displaystyle \int_{-B_r/2}^{B_r/2} S_s(f)\, df
    }{
    \displaystyle
    R_c
    \int_{-B_r/2}^{B_r/2} S_i(f)\, S_s(f)\, df
    }
    =
    \frac{
    \displaystyle \int_{-\infty}^{\infty} \left| H_R(f) \right|^2 S_s(f)\, df
    }{
    R_c \, \kappa_{is}
    },
\end{equation}
where $\kappa_{is}$ is referred to as the SSC~\cite{kaplan}, defined by\footnote{The SSC has units of seconds (or equivalently reciprocal hertz) and depends jointly on the spectral characteristics of the desired signal and the interference. Different interference signals may yield the same SSC with respect to a given desired signal; in such cases, they produce identical degradation in the effective carrier-to-noise density ratio, $(C_s/\!N_0)_{\mathrm{eff}}$, provided they are received with equal power. Consequently, if one interferer exhibits an SSC that is $x$~dB smaller than that of another interferer relative to the same desired signal, the two interferers will have the same impact on $(C_s/\!N_0)_{\mathrm{eff}}$ when the power of the former is increased by $x$~dB~\cite{kaplan}.}
\begin{equation}
\label{eq:ssc-def}
    \kappa_{is} = \int_{-B_r/2}^{B_r/2} S_i(f)\, S_s(f)\, df.
\end{equation}

\subsection{Receiver-Aware SSC and Performance Mapping}
\label{label_preliminaries_receiver}

While the preceding development follows the classical formulation of the SSC as presented in Kaplan and Hegarty~\cite{kaplan}, it is important to emphasize that the SSC is implicitly dependent on the receiver front-end and effective reception bandwidth. As clarified by Rügamer~\cite{ruegamer_dissertation}, the SSC is physically meaningful only over the frequency region passed by the receiver filter and may therefore be expressed by explicitly incorporating the receiver transfer function into the overlap integral as
\begin{equation}
\label{eq:ssc-receiver-aware}
    \kappa_{is}
    =
    \int_{-\infty}^{\infty}
    \left| H_R(f) \right|^2
    S_s(f)\,
    S_i(f)\,
    df,
\end{equation}
or equivalently by restricting the integration limits to the effective receiver bandwidth~\cite{kaplan}. In this interpretation, the SSC quantifies the spectral similarity between the desired signal and the interference after receiver filtering, rather than an abstract infinite-bandwidth property of the signals alone. Different receiver architectures or intermediate-frequency placements may yield different SSC values for the same pair of signals, reinforcing the role of the SSC as a receiver-dependent measure that directly governs interference-induced degradation of the effective carrier-to-noise density ratio~\cite{ruegamer_dissertation}.

For a fixed receiver architecture, the degradation induced by interference can be related to the spectral separation coefficient through the effective carrier-to-noise density ratio. Under standard correlator assumptions, this relationship may be expressed as
\begin{equation}
\label{eq:ssc-cn0-link}
    \left( \frac{C_s}{N_0} \right)_{\mathrm{eff}}
    =
    \frac{
    \int |H_R(f)|^2 S_s(f)\, df
    }{
    \displaystyle
    \frac{1}{(C_s/N_0)}
    +
    \left( \frac{C_i}{C_s} \right) \kappa_{is}
    },
\end{equation}
which highlights that, for a given receiver, interference impact is governed jointly by received interference power~\cite{kaplan}\footnote{Eq.~\ref{eq:ssc-cn0-link} shows that the SSC provides a direct mapping between interference and an equivalent degradation in carrier-to-noise density ratio for a fixed receiver architecture. The impact of interference on acquisition, carrier tracking, and demodulation can be characterized through the SSC without requiring explicit modeling of the underlying modulation-specific PSDs.}.
\section{Experiments}
\label{label_experiments}

For verification of the computed SSC and interference effects, we collect two distinct datasets: a real-world dataset acquired in an open-field environment using an arbitrary waveform  generator (AWG), and a simulated dataset generated using a Spirent GSS9000 RFCS.

\paragraph{Real-World Dataset} For the real-world dataset, the receiver is positioned at the center of an open field measuring $30\,\text{m} \times 50\,\text{m}$, while the AWG is placed at a distance of $20\,\text{m}$. The open field is surrounded by trees and small buildings, thereby introducing realistic multipath effects. Data are transmitted at a carrier frequency of $1.57542\,\text{GHz}$, at which the AWG transmits 210 interference instances with varying characteristics, covering the six general interference types: chirp, frequency hopper, modulated, pulsed, multitone, and noise. The interference bandwidth ranges from $0.1\,\text{MHz}$ to $60\,\text{MHz}$, and the signal power is set to $10\,\text{dBm}$. Further details on the AWG configurations are provided by Heublein et al.~\cite{heublein_feigl_crpa}. The receiver module with the wideband dipole whip antenna ANT-5GW-MMG2-SMA-1 records raw IQ samples over a bandwidth of $40.5\,\text{MHz}$, resulting in $122{,}880$ IQ samples on the L1 GPS band. The satellite signal power was assumed to be $-124\,\text{dBm}$, and the variable gain amplifier (VGA) was set to a value of $140$ ($\widehat{=}47\,\text{dB}$).

\paragraph{Simulated Dataset with RFCS} Using the RFCS, the satellite signal power was set to $-108\,\text{dBm}$. A VGA setting of either 146 ($\widehat{=}50\,\text{dB}$) or 160 ($\widehat{=}57\,\text{dB}$) was applied, and signals were generated in the E1 band using B/C signals. A total of 12 distinct IQ recordings were collected with the receiver module. Each recording consists of two consecutive segments: $1\,\text{min}$ without interference, followed by $1\,\text{min}$ with interference. The interference segment corresponds to one of the 12 variants listed in Table~\ref{table_results_real_world} (left part), comprising (1) two broad phase-shift keying (BPSK) matched spectrum interferences, (2) five noise interferences, and (3) five chirp interferences with a chirp rate of $5\,\text{kHz}$, both without and with frequency shift.

\paragraph{Receiver-Dependent Degradation Computation} A dual-band receiver module operating in the E1/E5 band is employed; however, only the E1 band is evaluated in this study. Data acquisition is performed with an $8\,\text{bit}$ IQ complex sampling directly to hard disc via an USB interface processed through a GNSS-SDR (allows the comparison between the real and computed $C\!/\!N_0$), an $8\,\text{bit}$ quantization depth, and a bandwidth and sampling rate of $40.5\,\text{MHz}$. The receiver has an analog RF-bandwidth of $60\,\text{MHz}$ but a digital filtering to $40.5\,\text{MHz}$

Next, we describe the computations explicitly for the considered receiver setup. The standard normalized PSD $S_s(f)$ of the GPS L1 or E1 B/C signal are used. The same is applied to the recorded signals in the presence of interference to obtain the normalized interference PSD $S_i(f)$. Subsequently, the SSC $\kappa_{is}$ is computed according to Eq.~\ref{eq:ssc-def}. The integral of the product $S_i(f) S_s(f)$ is evaluated numerically by approximating the integral as a step-wise summation over the effective receiver bandwidth $[-B_r/2, B_r/2]$. Using Eq.~\ref{eq:Q-def}, the jamming resistance quality factor is then obtained as $Q = \frac{1}{R_c \kappa_{is}}$, where $R_c = 1.023 \cdot 10^6\,\text{Hz}$ denotes the spreading code rate in chips per second. Next, the interference-induced degradation is computed using Eq.~\ref{eq:ssc-cn0-link}. For the simulated dataset, the interference power $C_i$ provided by the RFCS is used as reference (see Table~\ref{table_results_real_world}). For the real-world measurements, the interference power is estimated as
\begin{equation}
    C_i = P_{\text{sigcon}} - \left( 0.5 \cdot \text{VGA} - 23 \right)\,\text{dB}
\end{equation}
\vspace{-0.5cm}
\begin{equation}
    P_{\text{sigcon}} = 10 \cdot \log_{10} \Big( \frac{2(\sum_{i=1}^N \mathbf{U}_i^2)}{NR \cdot 1\,\text{mW} } \Big),
\end{equation}
where $\mathbf{U} = \frac{\sqrt{\mathbf{I}^2 + \mathbf{Q}^2}}{2^8}$ is the normalized magnitude (envelope) representation of in-phase component $\mathbf{I} = [I_1, I_2, \ldots, I_N]$ and quadrature component $\mathbf{Q} = [Q_1, Q_2, \ldots, Q_N]$, the division by $2^8$ serves as the normalization step, and $R = 50\,\Omega$.\footnote{The VGA is set to $140$ ($\widehat{=}47\,\text{dB}$) for the real-world data and to $146$ ($\widehat{=}50\,\text{dB}$) or $160$ ($\widehat{=}57\,\text{dB}$) for the simulated data. The received satellite signal power is computed and is $C_s = -124\,\text{dBm}$ for the real-world data and is $C_s = -108\,\text{dBm}$ for the simulated data. The noise power spectral density is $N_0 = -174\,\text{dBm/Hz}$ for the real-world data and $N_0 = -158\,\text{dBm/Hz}$ for the simulated data.} Using the computed SSC $\kappa_{is}$ from Eq.~\ref{eq:ssc-def}, the effective carrier-to-noise density ratio is obtained via Eq.~\ref{eq:ssc-cn0-link}. Finally, the interference-induced degradation is expressed as
\begin{equation}
    \Delta \approx \left( \frac{C_s}{N_0} \right) - \left( \frac{C_s}{N_0} \right)_{\mathrm{eff}}.
\end{equation}

\section{Evaluation}
\label{label_evaluation}

\setlength{\intextsep}{6pt}
\setlength{\columnsep}{12pt}
\begin{wrapfigure}{R}{3.9cm}
    \begin{minipage}[b]{1.0\linewidth}
        \centering
        \vspace{-0.15cm}
        \includegraphics[trim=10 10 10 10, clip, width=1.0\linewidth]{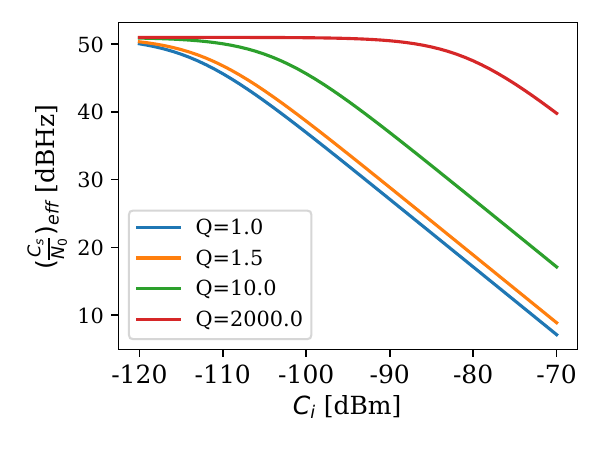}
        \vspace{-0.5cm}
        \caption{Correlation between $(C_s/\!N_0)_{\text{eff}}$ and $C_{i}$ dependent on $Q$.}
    \label{figure_evaluation_Q_power}
    \end{minipage}
\end{wrapfigure}
\paragraph{Ablation: Influence of $Q$ on $(C_s/N_0)_{\text{eff}}$ vs.\ $C_i$} To isolate the role of the jamming resistance quality factor $Q$, we perform a short ablation in which all receiver and signal parameters are held constant\footnote{$C_s = -122.9\,\text{dBm}$, $N_0 = 10\cdot \log_{10} (TK_b) = -173.9\,\text{dBm}$, $T=290K$} and only $Q$ is varied. For each fixed $Q$, we evaluate the effective carrier-to-noise density ratio $(C_s/\!N_0)_{\text{eff}}$ as a function of the interference input power $C_i$ (in dBm). The resulting curves are shown in Fig.~\ref{figure_evaluation_Q_power}, which shows a clear monotonic relationship between $C_{i}$ and $(C_s/\!N_0)_{\text{eff}}$: Increasing interference

\newcommand\x{-0.5cm}
\begin{figure*}[!t]
    \centering
	\begin{minipage}[t]{0.325\linewidth}
        \centering
    	\includegraphics[trim=10 10 10 10, clip, width=1.0\linewidth]{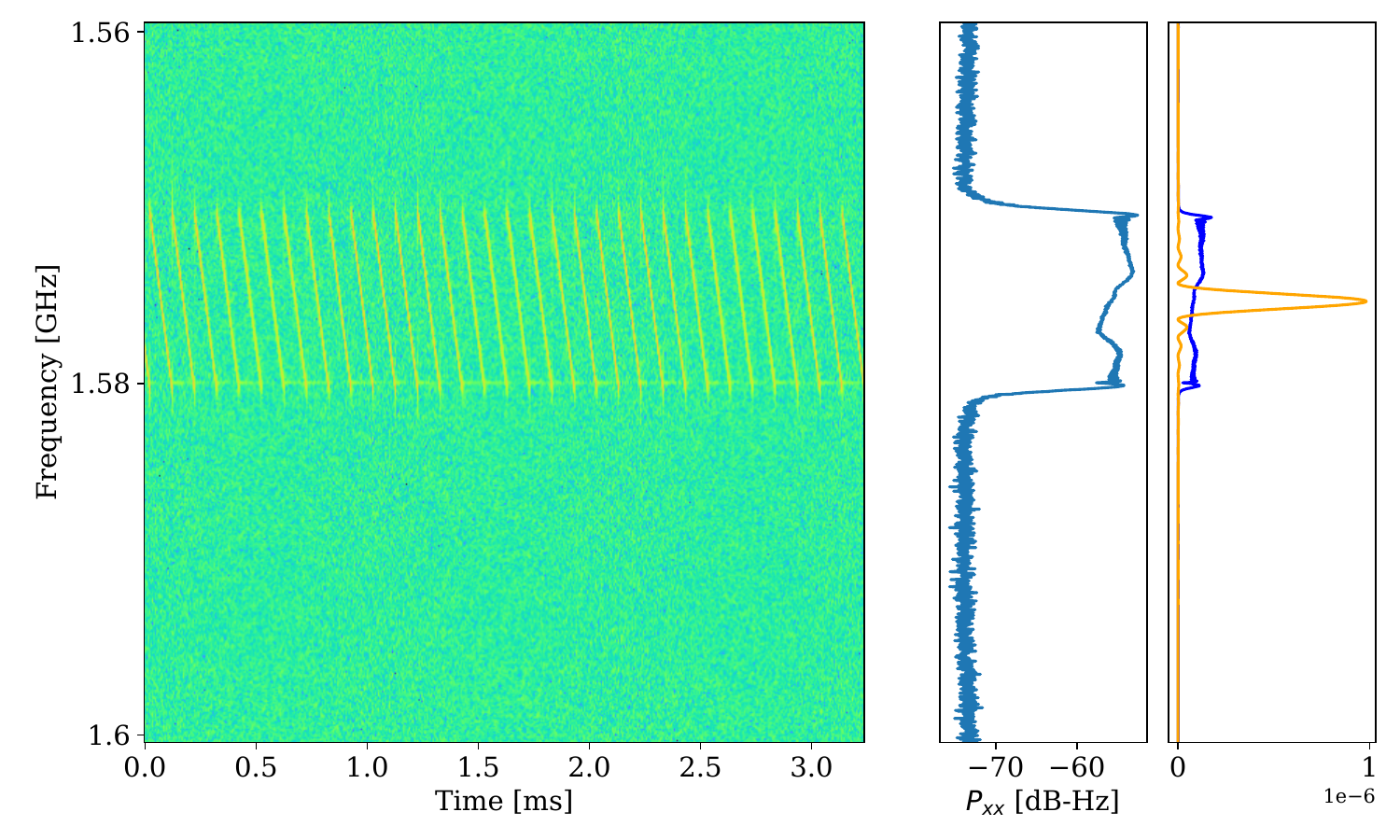}
        \vspace{\x}
        \subcaption{Chirp, $10\,\text{MHz}$ BW, $Q = 12.59$, $C_i = -63.81\,\text{dBm}$, $\Delta = 39.07\,\text{dB}$.}
        \label{figure_evaluation_real_simulated1}
    \end{minipage}
    \hfill
	\begin{minipage}[t]{0.325\linewidth}
        \centering
    	\includegraphics[trim=10 10 10 10, clip, width=1.0\linewidth]{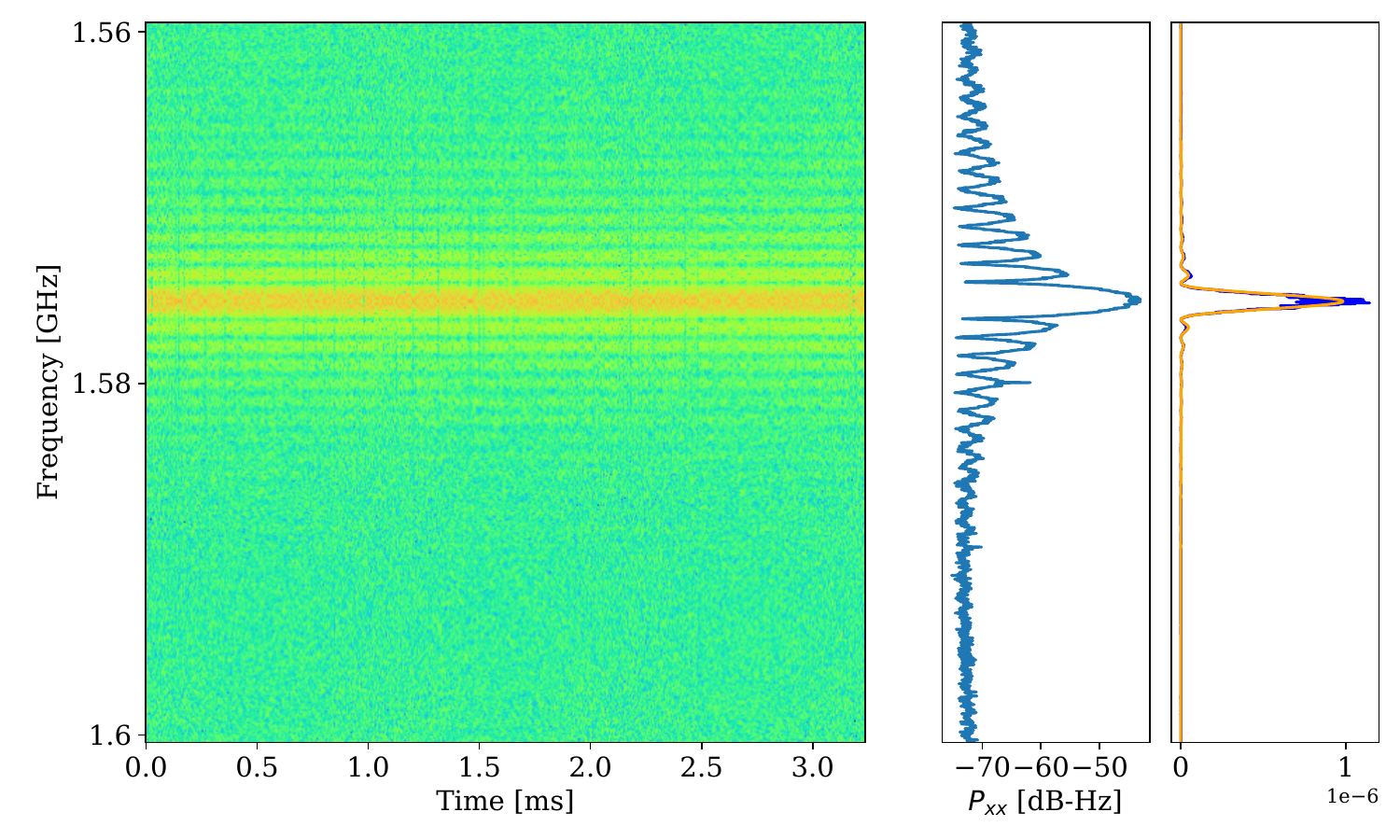}
        \vspace{\x}
        \subcaption{BPSK, $1\,\text{MHz}$ BW main lobe, $Q = 1.54$, $C_i = -62.70\,\text{dBm}$, $\Delta = 49.30\,\text{dB}$.}
        \label{figure_evaluation_real_simulated2}
    \end{minipage}
    \hfill
	\begin{minipage}[t]{0.325\linewidth}
        \centering
    	\includegraphics[trim=10 10 10 10, clip, width=1.0\linewidth]{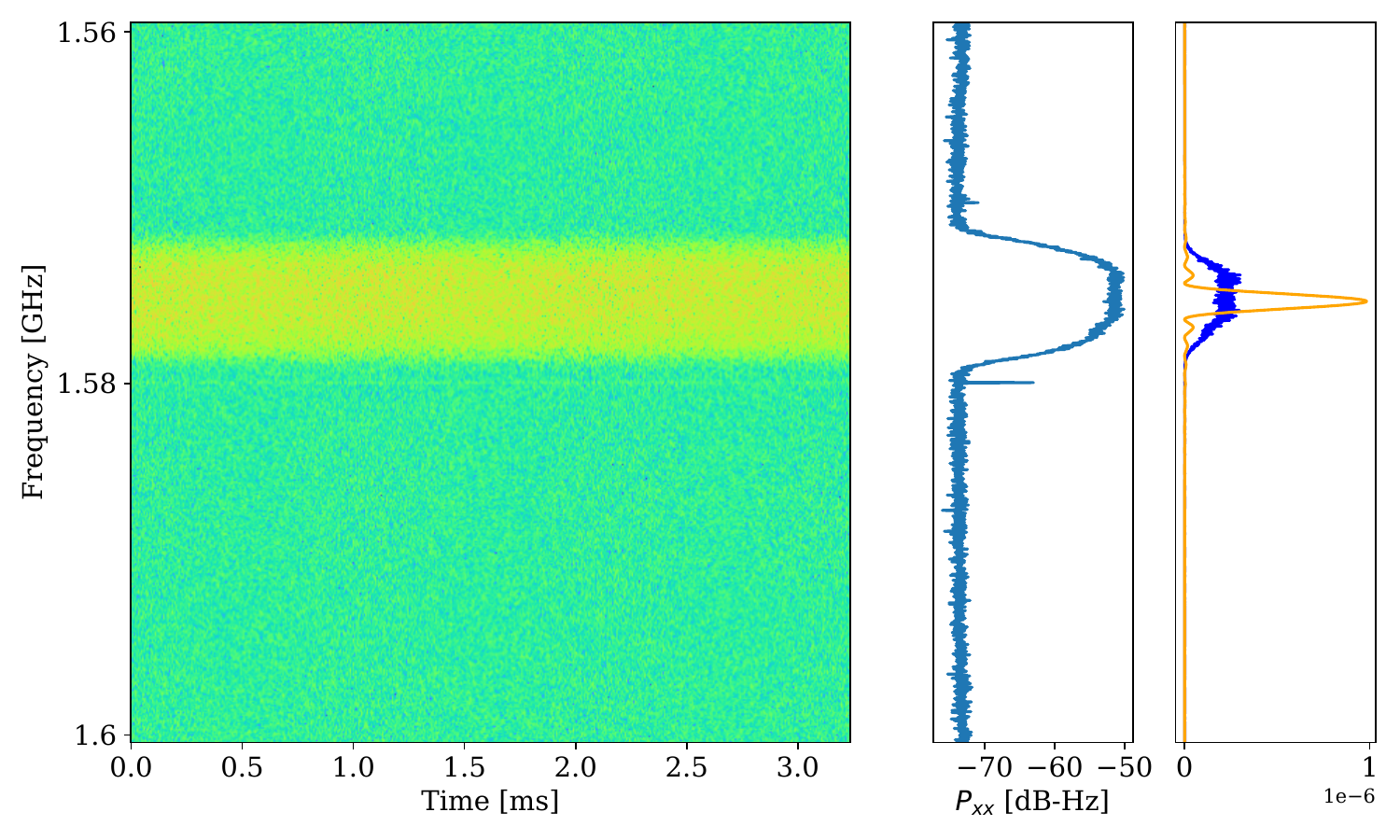}
        \vspace{\x}
        \subcaption{Noise, $5\,\text{MHz}$ BW, $Q = 4.60$, $C_i = -63.92\,\text{dBm}$, $\Delta = 43.34\,\text{dB}$.}
        \label{figure_evaluation_real_simulated3}
    \end{minipage}
	\begin{minipage}[t]{0.325\linewidth}
        \centering
    	\includegraphics[trim=10 10 10 10, clip, width=1.0\linewidth]{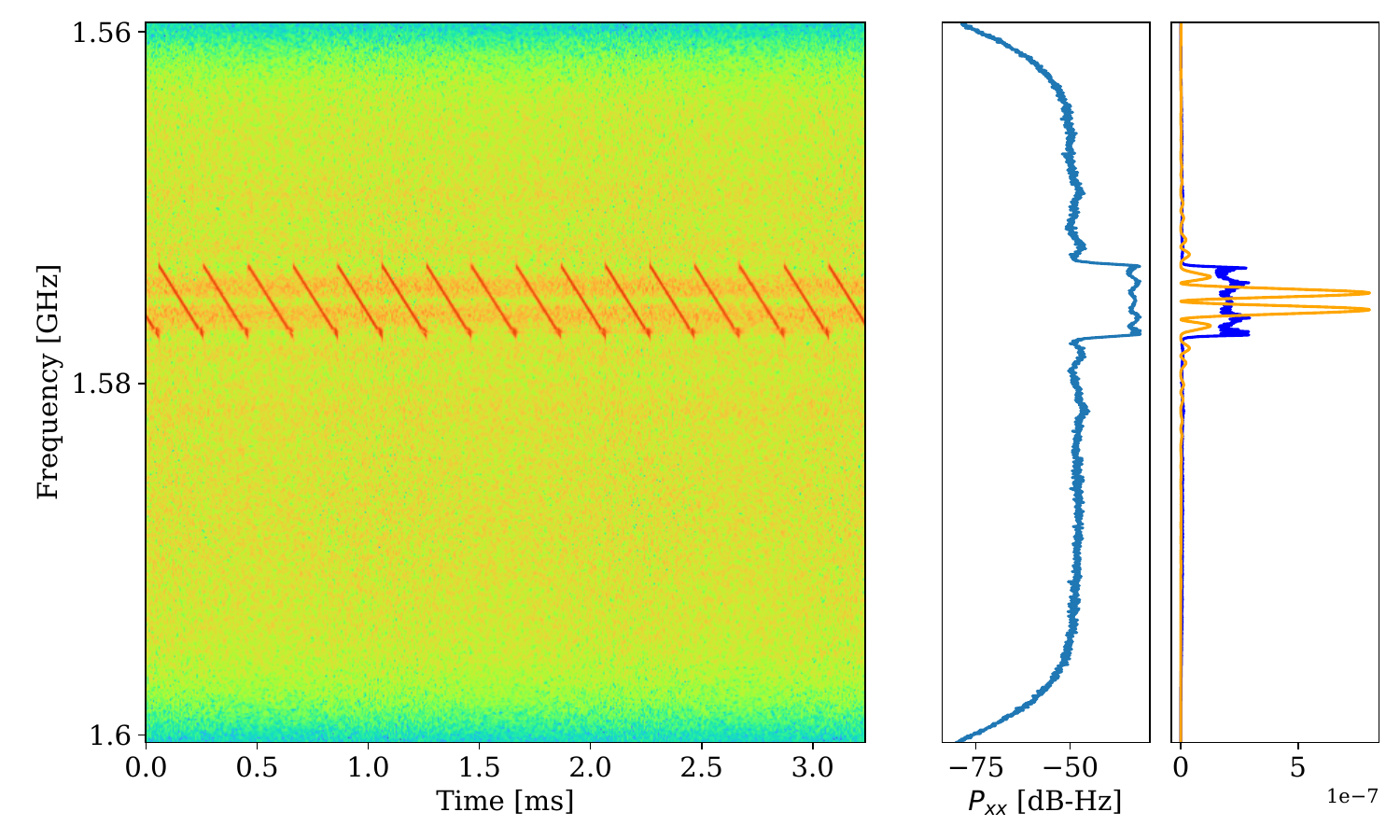}
        \vspace{\x}
        \subcaption{Chirp, $4\,\text{MHz}$ BW, $Q = 5.37$, $C_i = -84.51\,\text{dBm}$, $\Delta_{C_i} = 6.26\,\text{dB}$, $\Delta_{C_{i,\text{ref}}} = 6.66\,\text{dB}$, $\Delta_{C_{i,\text{meas}}} = 5.8\,\text{dB}$.}
        \label{figure_evaluation_real_simulated4}
    \end{minipage}
    \hfill
	\begin{minipage}[t]{0.325\linewidth}
        \centering
    	\includegraphics[trim=10 10 10 10, clip, width=1.0\linewidth]{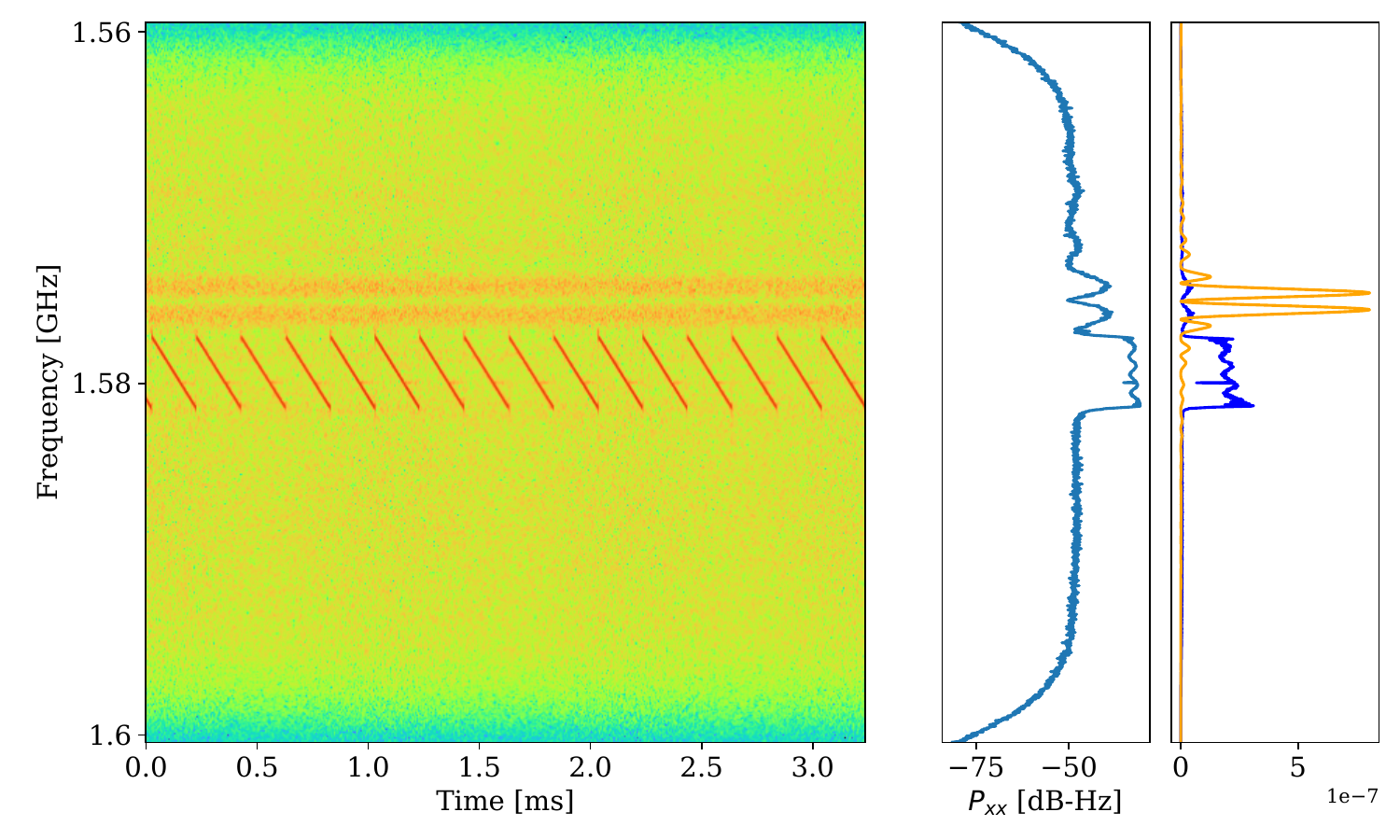}
        \vspace{\x}
        \subcaption{Chirp, $4\,\text{MHz}$ BW, $4\,\text{MHz}$ shift, $Q = 42.37$, $C_i = -83.55\,\text{dBm}$, $\Delta_{C_i} = 1.79\,\text{dB}$, $\Delta_{C_{i,\text{ref}}} = 1.64\,\text{dB}$, $\Delta_{C_{i,\text{meas}}} = 1.4\,\text{dB}$.}
        \label{figure_evaluation_real_simulated5}
    \end{minipage}
    \hfill
	\begin{minipage}[t]{0.325\linewidth}
        \centering
    	\includegraphics[trim=10 10 10 10, clip, width=1.0\linewidth]{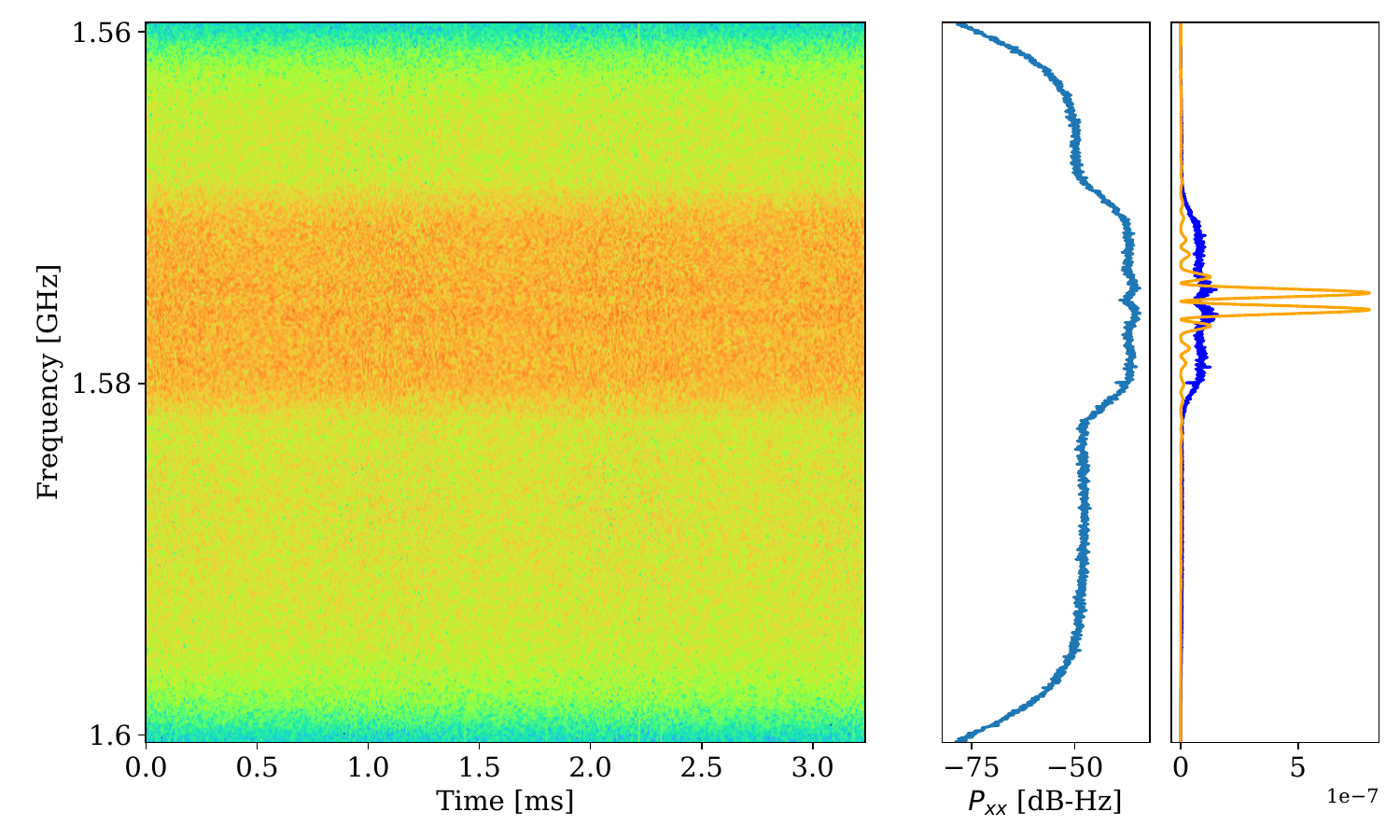}
        \vspace{\x}
        \subcaption{Noise, $10\,\text{MHz}$ BW, $Q = 10.85$, $C_i = -84.13\,\text{dBm}$, $\Delta_{C_i} = 4.39\,\text{dB}$, $\Delta_{C_{i,\text{ref}}} = 4.47\,\text{dB}$, $\Delta_{C_{i,\text{meas}}} = 4.4\,\text{dB}$.}
        \label{figure_evaluation_real_simulated6}
    \end{minipage}
    \vspace{-0.1cm}
    \caption{Presentation of signal+interference spectrograms (left), and the Welch PSD (middle) with normalized signal (right; for the standard satellite signal in orange; for the received signal in blue) for the real-world (top, GPS L1) and the simulated (bottom, E1 B/C) datasets. $Q$: jamming resistance quality factor, $\Delta_{C_i}$: degradation dependent on computed interference power $C_i$, $\Delta_{C_{i,\text{ref}}}$: degradation dependent on the reference signal power of Spirent, $\Delta_{C_{i,\text{meas}}}$: degradation measured by post-processing.}
    \label{figure_evaluation_real_simulated}
    \vspace{-0.2cm}
\end{figure*}

\noindent power (i.e., $C_i$ becomes less negative) yields a pronounced reduction in $(C_s/\!N_0)_{\text{eff}}$ for all considered $Q$. Importantly, the rate of degradation is strongly controlled by $Q$. For low values ($Q \in \{1.0,1.5\}$), $(C_s/\!N_0)_{\text{eff}}$ decreases steeply across the full $C_{i}$ range, indicating high susceptibility to interference. Increasing $Q$ to $10$ substantially mitigates this loss, shifting the curve and reducing the sensitivity to $C_i$. For $Q = 2{,}000$, $(C_s/\!N_0)_{\text{eff}}$ remains close to its interference-free level over a wide range of $C_i$ and only degrades noticeably at the highest interference powers, confirming that larger $Q$ (equivalently smaller SSC $\kappa_{is}$) corresponds to increased jamming resistance and improved robustness of the receiver against interference.

\begin{figure}[!t]
    \centering
    \includegraphics[trim=11 11 11 11, clip, width=0.9\linewidth]{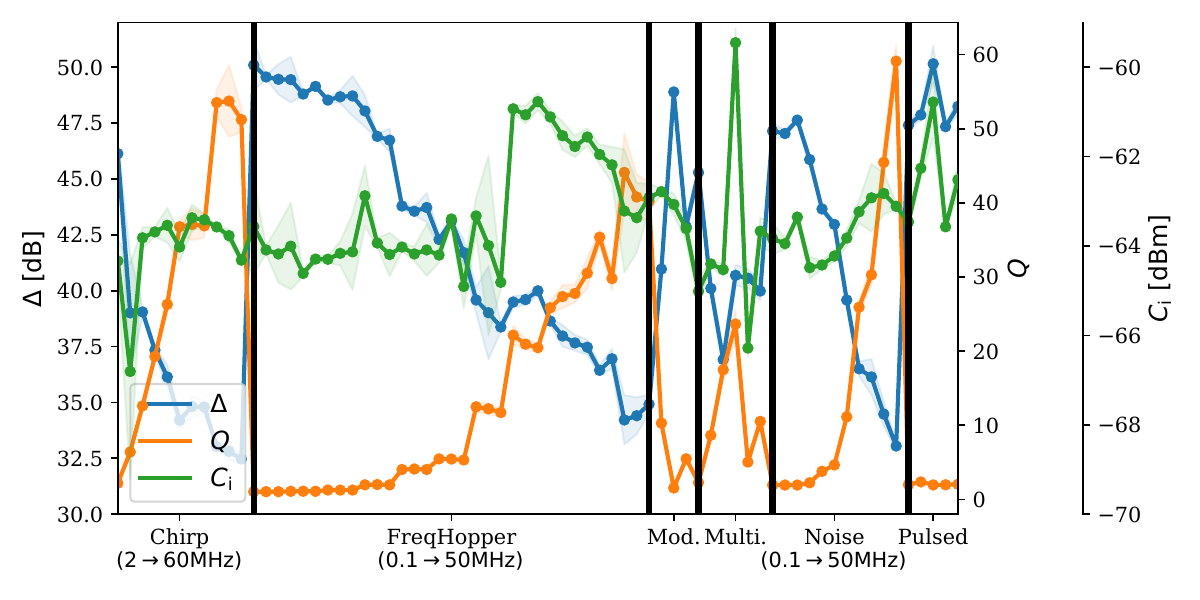}
    \vspace{-0.2cm}
    \caption{Overview of jamming resistance $Q$, interference power $C_i$, and degradation $\Delta$ for the real-world dataset, averaged over $\approx 1{,}600$ IQ interference samples with standard deviation.}
    \label{figure_results_interferences_real_world}
    \vspace{-0.25cm}
\end{figure}

\begin{figure*}[!t]
    \centering
	\begin{minipage}[t]{0.245\linewidth}
        \centering
    	\includegraphics[trim=11 11 11 11, clip, width=1.0\linewidth]{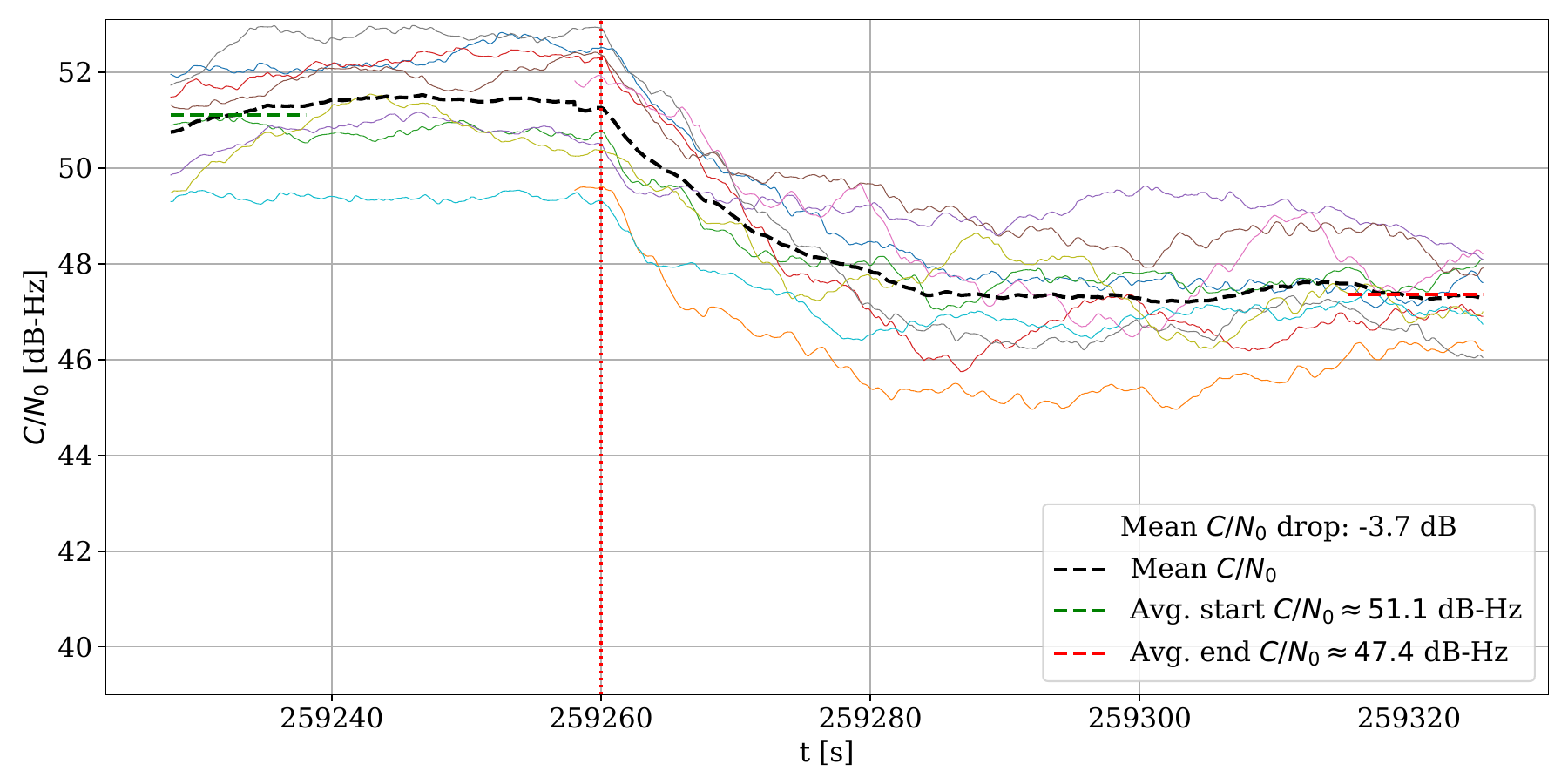}
        \subcaption{Modulated BPSK, $-84\,\text{dBm}$.}
        \label{figure_plot_CN1}
    \end{minipage}
    \hfill
	\begin{minipage}[t]{0.245\linewidth}
        \centering
    	\includegraphics[trim=11 11 11 11, clip, width=1.0\linewidth]{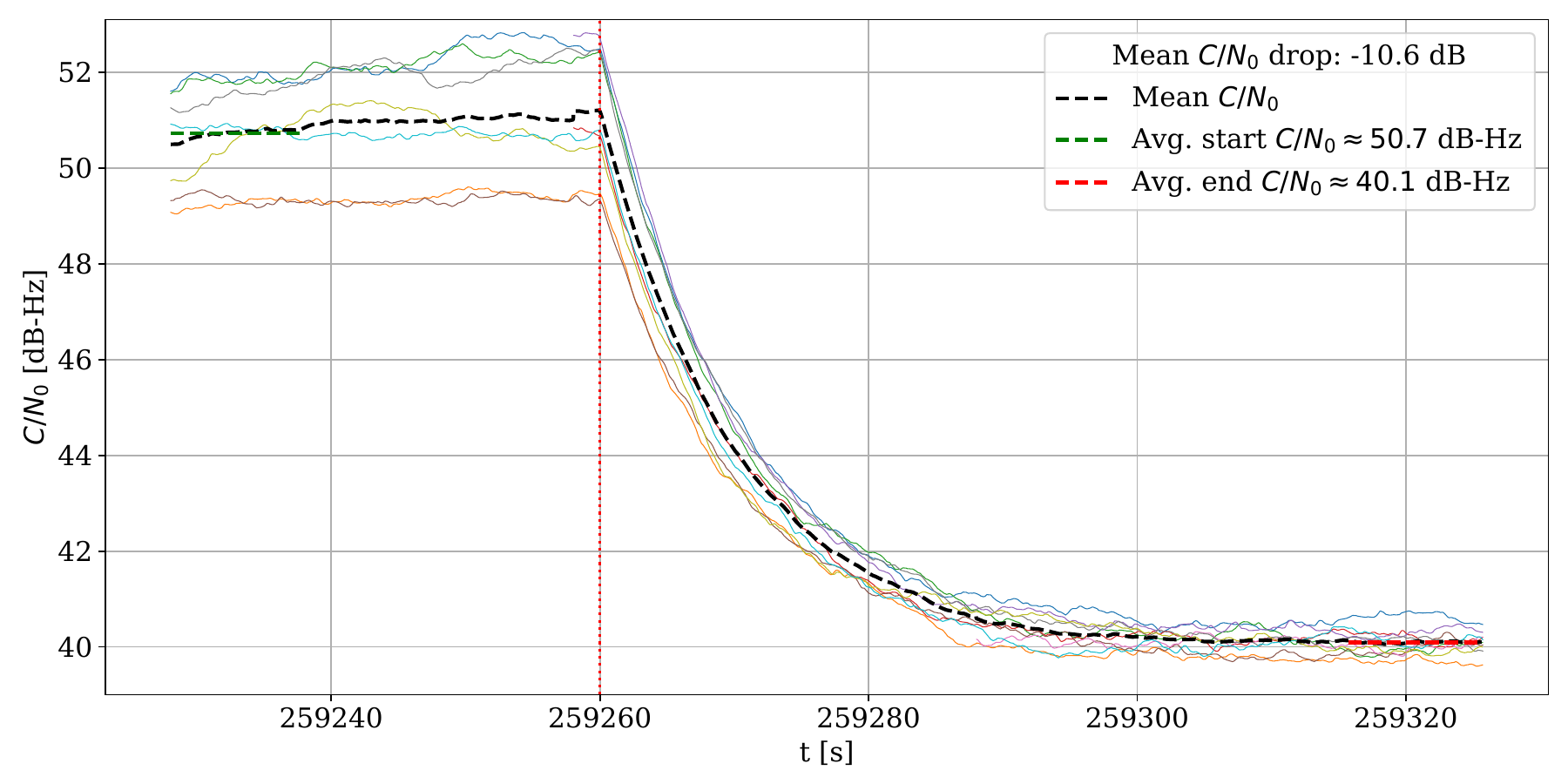}
        \subcaption{Noise, $-74\,\text{dBm}$, $20\,\text{MHz}$ BW.}
        \label{figure_plot_CN2}
    \end{minipage}
    \hfill
	\begin{minipage}[t]{0.245\linewidth}
        \centering
    	\includegraphics[trim=11 11 11 11, clip, width=1.0\linewidth]{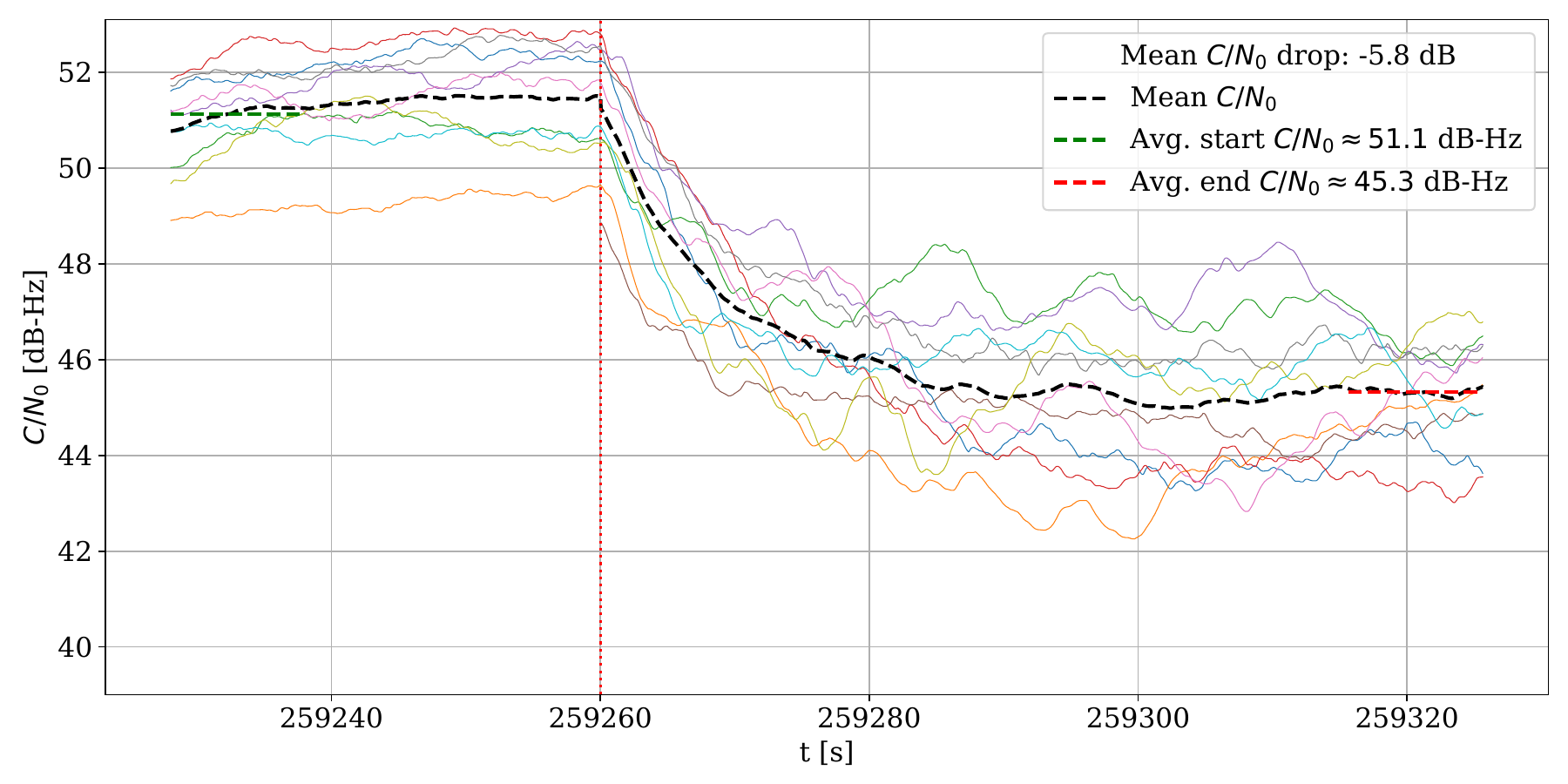}
        \subcaption{Chirp, $-84\,\text{dBm}$, $4\,\text{MHz}$ BW.}
        \label{figure_plot_CN3}
    \end{minipage}
    \hfill
	\begin{minipage}[t]{0.245\linewidth}
        \centering
    	\includegraphics[trim=11 11 11 11, clip, width=1.0\linewidth]{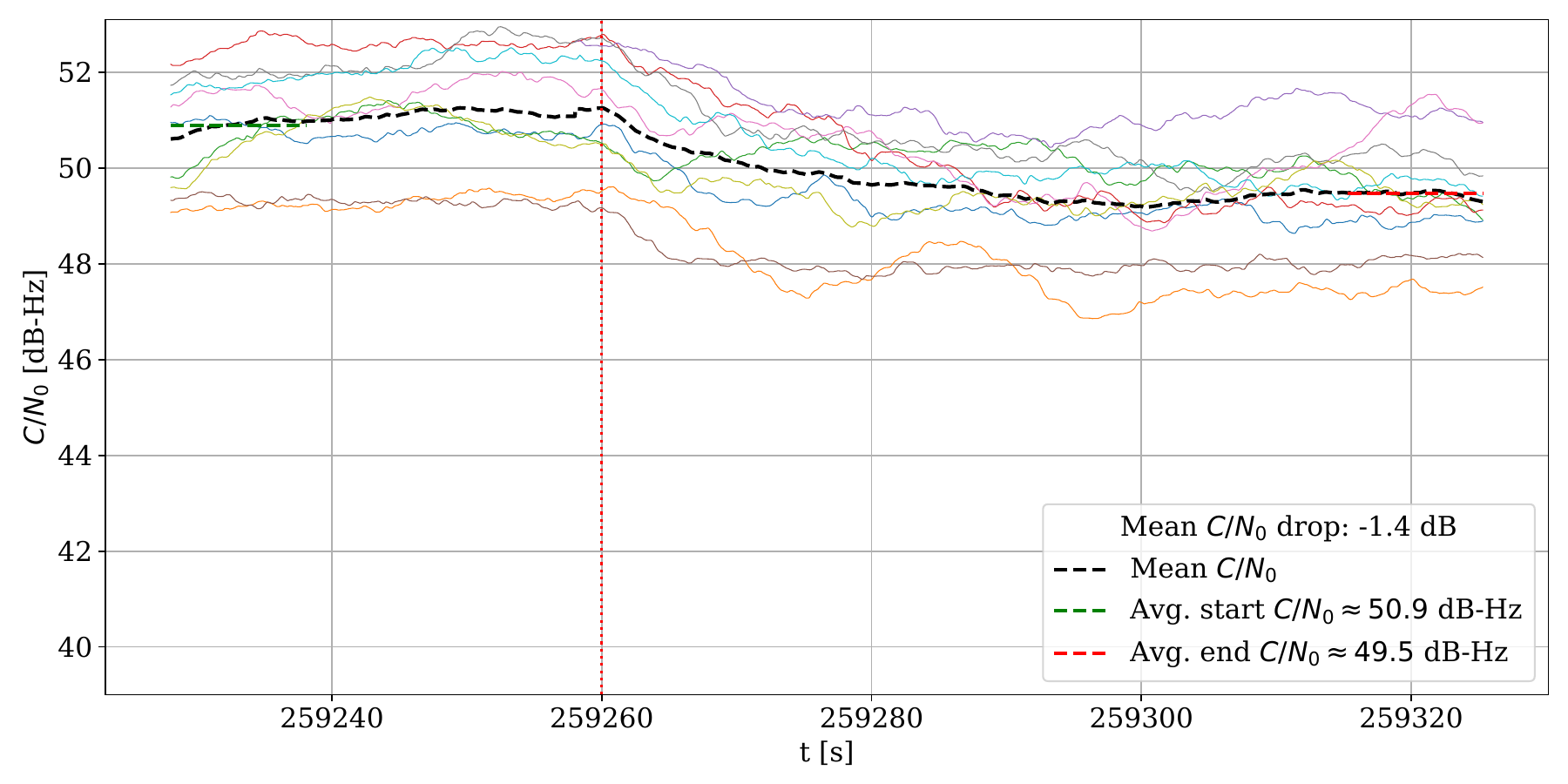}
        \subcaption{Chirp, $-84\,\text{dBm}$, $4\,\text{MHz}$ shift.}
        \label{figure_plot_CN4}
    \end{minipage}
    \vspace{-0.15cm}
    \caption{Plot of $C\!/\!N_0$ loss for each visible satellite for four different interferences (first period shows no interference).}
    \label{figure_plot_CN}
\end{figure*}

\renewcommand{\arraystretch}{0.9}
\begin{table*}[t!]
\begin{center}
\setlength{\tabcolsep}{2.8pt}
    \caption{Overview of interference recordings with Spirent (left part) and the corresponding jamming resistance $Q$, interference power (either computed, $C_i$, or with reference from Spirent, $C_\text{i,ref}$), and degradation $\Delta$ dependent on $C_i$ (right part) of the simulated dataset. Results are averaged over 10 IQ samples and we present standard deviation ($\pm$).}
    \label{table_results_real_world}
    \vspace{-0.1cm}
    \begin{tabular}{ p{0.5cm} | p{0.5cm} | p{0.5cm} | p{0.5cm} | p{0.5cm} | p{0.5cm} | p{0.5cm} | p{0.5cm} | p{0.5cm} | p{0.5cm} }
    \multicolumn{1}{c}{\textbf{Interference}} & \multicolumn{1}{c}{\textbf{Amp.}} & \multicolumn{1}{c}{\textbf{Bandwidth (BW)}} & \multicolumn{1}{c||}{\textbf{Frequency Shift}} & \multicolumn{1}{c}{$Q$ [$\cdot$]} & \multicolumn{1}{c}{$C_i$ [dBm]} & \multicolumn{1}{c}{$C_{i,\text{ref}}$ [dBm]} & \multicolumn{1}{c}{$\Delta_{C_i}$ [dB]} & \multicolumn{1}{c}{$\Delta_{C_{i,\text{ref}}}$ [dB]} & \multicolumn{1}{c}{$\Delta_{C_{i,\text{meas}}}$ [dB]} \\ \hline
    \multicolumn{1}{l}{BPSK} & \multicolumn{1}{r}{$57\,\text{dB}$} & \multicolumn{1}{r}{$\approx10\,\text{MHz}$} & \multicolumn{1}{r||}{$0\,\text{MHz}$} & \multicolumn{1}{r}{$10.92\,\pm\,0.08$} & \multicolumn{1}{r}{$-84.50\,\pm\,0.01$} & \multicolumn{1}{r}{$-84.0$} & \multicolumn{1}{r}{$4.14\,\pm\,0.02$} & \multicolumn{1}{r}{$4.45\,\pm\,0.02$} & \multicolumn{1}{r}{$3.7$} \\
    \multicolumn{1}{l}{BPSK} & \multicolumn{1}{r}{$57\,\text{dB}$} & \multicolumn{1}{r}{$\approx10\,\text{MHz}$} & \multicolumn{1}{r||}{$0\,\text{MHz}$} & \multicolumn{1}{r}{$9.70\,\pm\,0.15$} & \multicolumn{1}{r}{$-91.83\,\pm\,0.02$} & \multicolumn{1}{r}{$-94.0$} & \multicolumn{1}{r}{$1.24\,\pm\,0.02$} & \multicolumn{1}{r}{$0.80\,\pm\,0.01$} & \multicolumn{1}{r}{$0.4$} \\
    \multicolumn{1}{l}{Noise} & \multicolumn{1}{r}{$57\,\text{dB}$} & \multicolumn{1}{r}{$10\,\text{MHz}$} & \multicolumn{1}{r||}{$0\,\text{MHz}$} & \multicolumn{1}{r}{$10.89\,\pm\,0.13$} & \multicolumn{1}{r}{$-84.15\,\pm\,0.02$} & \multicolumn{1}{r}{$-84.0$} & \multicolumn{1}{r}{$4.36\,\pm\,0.03$} & \multicolumn{1}{r}{$4.46\,\pm\,0.03$} & \multicolumn{1}{r}{$4.4$} \\
    \multicolumn{1}{l}{Noise} & \multicolumn{1}{r}{$57\,\text{dB}$} & \multicolumn{1}{r}{$10\,\text{MHz}$} & \multicolumn{1}{r||}{$0\,\text{MHz}$} & \multicolumn{1}{r}{$9.70\,\pm\,0.09$} & \multicolumn{1}{r}{$-91.68\,\pm\,0.03$} & \multicolumn{1}{r}{$-94.0$} & \multicolumn{1}{r}{$1.28\,\pm\,0.01$} & \multicolumn{1}{r}{$0.80\,\pm\,0.01$} & \multicolumn{1}{r}{0.4} \\
    \multicolumn{1}{l}{Noise} & \multicolumn{1}{r}{$57\,\text{dB}$} & \multicolumn{1}{r}{$20\,\text{MHz}$} & \multicolumn{1}{r||}{$0\,\text{MHz}$} & \multicolumn{1}{r}{$24.60\,\pm\,0.34$} & \multicolumn{1}{r}{$-73.07\,\pm\,0.01$} & \multicolumn{1}{r}{$-74.0$} & \multicolumn{1}{r}{$10.34\,\pm\,0.05$} & \multicolumn{1}{r}{$9.51\,\pm\,0.05$} & \multicolumn{1}{r}{$11.1$} \\
    \multicolumn{1}{l}{Noise} & \multicolumn{1}{r}{$50\,\text{dB}$} & \multicolumn{1}{r}{$20\,\text{MHz}$} & \multicolumn{1}{r||}{$0\,\text{MHz}$} & \multicolumn{1}{r}{$24.55\,\pm\,0.20$} & \multicolumn{1}{r}{$-71.60\,\pm\,0.02$} & \multicolumn{1}{r}{$-74.0$} & \multicolumn{1}{r}{$11.70\,\pm\,0.04$} & \multicolumn{1}{r}{$9.52\,\pm\,0.03$} & \multicolumn{1}{r}{$10.6$} \\
    \multicolumn{1}{l}{Noise} & \multicolumn{1}{r}{$57\,\text{dB}$} & \multicolumn{1}{r}{$20\,\text{MHz}$} & \multicolumn{1}{r||}{$0\,\text{MHz}$} & \multicolumn{1}{r}{$18.20\,\pm\,0.22$} & \multicolumn{1}{r}{$-83.77\,\pm\,0.02$} & \multicolumn{1}{r}{$-84.0$} & \multicolumn{1}{r}{$3.29\,\pm\,0.03$} & \multicolumn{1}{r}{$3.16\,\pm\,0.03$} & \multicolumn{1}{r}{$3.0$} \\
    \multicolumn{1}{l}{Chirp, $5\,\text{kHz}$} & \multicolumn{1}{r}{$57\,\text{dB}$} & \multicolumn{1}{r}{$4\,\text{MHz}$} & \multicolumn{1}{r||}{$0\,\text{MHz}$} & \multicolumn{1}{r}{$5.10\,\pm\,0.17$} & \multicolumn{1}{r}{$-84.48\,\pm\,0.03$} & \multicolumn{1}{r}{$-84.0$} & \multicolumn{1}{r}{$6.46\,\pm\,0.12$} & \multicolumn{1}{r}{$6.83\,\pm\,0.11$} & \multicolumn{1}{r}{$5.8$} \\
    \multicolumn{1}{l}{Chirp, $5\,\text{kHz}$} & \multicolumn{1}{r}{$57\,\text{dB}$} & \multicolumn{1}{r}{$4\,\text{MHz}$} & \multicolumn{1}{r||}{$0\,\text{MHz}$} & \multicolumn{1}{r}{$7.41\,\pm\,0.18$} & \multicolumn{1}{r}{$-91.82\,\pm\,0.03$} & \multicolumn{1}{r}{$-94.0$} & \multicolumn{1}{r}{$1.57\,\pm\,0.04$} & \multicolumn{1}{r}{$1.02\,\pm\,0.02$} & \multicolumn{1}{r}{$0.5$} \\
    \multicolumn{1}{l}{Chirp, $5\,\text{kHz}$} & \multicolumn{1}{r}{$57\,\text{dB}$} & \multicolumn{1}{r}{$20\,\text{MHz}$} & \multicolumn{1}{r||}{$0\,\text{MHz}$} & \multicolumn{1}{r}{$18.17\,\pm\,0.53$} & \multicolumn{1}{r}{$-83.81\,\pm\,0.02$} & \multicolumn{1}{r}{$-84.0$} & \multicolumn{1}{r}{$3.27\,\pm\,0.07$} & \multicolumn{1}{r}{$3.17\,\pm\,0.07$} & \multicolumn{1}{r}{$2.9$} \\
    \multicolumn{1}{l}{Chirp, $5\,\text{kHz}$} & \multicolumn{1}{r}{$57\,\text{dB}$} & \multicolumn{1}{r}{$20\,\text{MHz}$} & \multicolumn{1}{r||}{$0\,\text{MHz}$} & \multicolumn{1}{r}{$11.06\,\pm\,0.12$} & \multicolumn{1}{r}{$-91.62\,\pm\,0.03$} & \multicolumn{1}{r}{$-94.0$} & \multicolumn{1}{r}{$1.16\,\pm\,0.01$} & \multicolumn{1}{r}{$0.71\,\pm\,0.01$} & \multicolumn{1}{r}{$0.2$} \\
    \multicolumn{1}{l}{Chirp, $5\,\text{kHz}$} & \multicolumn{1}{r}{$57\,\text{dB}$} & \multicolumn{1}{r}{$4\,\text{MHz}$} & \multicolumn{1}{r||}{$4\,\text{MHz}$} & \multicolumn{1}{r}{$42.42\,\pm\,0.66$} & \multicolumn{1}{r}{$-83.56\,\pm\,0.01$} & \multicolumn{1}{r}{$-84.0$} & \multicolumn{1}{r}{$1.79\,\pm\,0.02$} & \multicolumn{1}{r}{$1.64\,\pm\,0.02$} & \multicolumn{1}{r}{$1.4$} \\
    \end{tabular}
    \vspace{-0.4cm}
\end{center}
\end{table*}

\paragraph{Evaluation Results on the Real-World Dataset} Fig.~\ref{figure_results_interferences_real_world} summarizes the evaluation results obtained from the real-world interference dataset comprising multiple interference types with diverse spectral characteristics. The jamming resistance $Q$ exhibits a clear relationship with both the interference power $C_i$ and the observed degradation $\Delta$ across all considered scenarios. Interferences exhibiting stronger spectral overlap with the desired satellite signal and higher spectral concentration within the receiver passband result in increased degradation. Conversely, increasing the interference BW generally leads to higher $Q$ values and a corresponding reduction in degradation $\Delta$. Fig.~\ref{figure_evaluation_real_simulated1} to \ref{figure_evaluation_real_simulated3} compares three representative interference types by jointly analyzing their time-frequency structure, spectral characteristics, and resulting performance impact. The wideband chirp interference in Fig.~\ref{figure_evaluation_real_simulated1} exhibits a large occupied BW and pronounced spectral overlap with the satellite signal, resulting in a relatively low jamming resistance quality factor ($Q = 12.59$) and a substantial degradation of $\Delta = 39.07\,\text{dB}$ despite moderate interference power. In Fig.~\ref{figure_evaluation_real_simulated2}, the modulated BPSK interference is spectrally concentrated within the main lobe of the desired signal, yielding a significantly smaller $Q = 1.54$ and the largest computed degradation ($\Delta = 49.30\,\text{dB}$), even though the interference power is comparable. Conversely, the wideband noise interference in Fig.~\ref{figure_evaluation_real_simulated3} spans a larger BW but exhibits a more uniform spectral distribution, leading to a higher $Q = 4.60$ and a reduced degradation of $\Delta = 43.34\,\text{dB}$ relative to the BPSK case. Overall, the results demonstrate that interference-induced degradation is governed not solely by interference power $C_i$, but predominantly by spectral overlap as quantified by the jamming resistance quality factor $Q$. Interferences with stronger spectral alignment to the desired signal yield lower $Q$ values and disproportionately higher $C\!/\!N_0$ degradation, confirming the suitability of $Q$ as a compact and physically meaningful metric for assessing interference impact.

\paragraph{Evaluation Results on the Simulated Dataset} Fig.~\ref{figure_plot_CN} shows the evolution of the $C\!/\!N_0$ ratio for all visible satellites under four representative interference scenarios. Prior to interference activation, all satellites exhibit stable $C\!/\!N_0$ levels. The BPSK interference in Fig.~\ref{figure_plot_CN1} causes a moderate degradation, indicating partial spectral overlap with the desired signal. In contrast, the wideband noise interference in Fig.~\ref{figure_plot_CN2} leads to an abrupt and nearly uniform $C\!/\!N_0$ collapse across all satellites, reflecting strong spectral overlap over the entire receiver BW. The narrowband chirp interference in Fig.~\ref{figure_plot_CN3} results in a smaller and more heterogeneous degradation, while the frequency-shifted chirp in Fig.~\ref{figure_plot_CN4} exhibits the weakest impact, confirming that reduced spectral overlap mitigates interference-induced $C\!/\!N_0$ loss. The $C\!/\!N_0$ does not have a direct analytical relationship with tracking accuracy~\cite{betz_kolodziejski}.

Table~\ref{table_results_real_world} shows a clear relationship between the jamming resistance quality factor $Q$ and the observed degradation $\Delta$. For a given interference type, changes in the interference power $C_i$ primarily affect the magnitude of the degradation, while the corresponding $Q$ values remain largely invariant, indicating that $Q$ is predominantly determined by the spectral characteristics of the interference rather than by its absolute power. Interference signals with increased BW or frequency offsets yield higher $Q$ values and, consequently, reduced degradation even at comparable power levels. Furthermore, slightly lower $Q$ values are observed for weaker interference levels (e.g., $C_i = -91.83\,\text{dBm}$) compared to stronger interference cases, as the desired satellite signal dominates the received spectrum at low interference power, resulting in increased normalized spectral overlap. Overall, these results confirm that $Q$ constitutes a compact, robust, and receiver-aware metric for predicting interference-induced performance degradation.

Fig.~\ref{figure_evaluation_real_simulated4} to \ref{figure_evaluation_real_simulated6} evaluates simulated interference scenarios and the sensitivity of the jamming resistance quality factor $Q$ to BW and frequency shift. Conversely, the narrowband chirp in Fig.~\ref{figure_evaluation_real_simulated4} overlaps more strongly with the signal spectrum, leading to a lower $Q = 5.37$ and an increased degradation of $\Delta_{C_{i,\text{ref}}} = 6.66\,\text{dB}$. The frequency-shifted chirp in Fig.~\ref{figure_evaluation_real_simulated5} exhibits limited overlap with the desired signal, resulting in a comparatively higher $Q = 42.37$ and a moderate degradation of $\Delta_{C_{i,\text{ref}}} = 1.64\,\text{dB}$. In Fig.~\ref{figure_evaluation_real_simulated6}, the wideband noise spans a larger BW but remains spectrally less aligned with the signal spectrum, yielding a lower jamming resistance quality factor ($Q = 10.85$) and only minor degradation ($\Delta_{C_{i,\text{ref}}} = 4.47\,\text{dB}$). These results further confirm that interference-induced degradation is not determined by interference power alone but is primarily governed by spectral overlap as captured by $Q$. Interference signals with reduced spectral alignment achieve substantially higher $Q$ values and correspondingly lower performance degradation, reinforcing the validity of the jamming resistance quality factor as a receiver-aware metric for interference impact assessment.
\section{Conclusion}
\label{label_conclusion}

\blfootnote{\textbf{Acknowledgments.} We thank Frank Förster and Simon Kocher for their support with receiver configuration and dataset generation. This work has been carried out within the DARCII project, funding code 50NA2401, supported by the German Federal Ministry for Economic Affairs and Climate Action (BMWK), managed by the German Space Agency at DLR and assisted by the Bundesnetzagentur (BNetzA) and the Federal Agency for Cartography and Geodesy (BKG).}
We used a receiver-aware SSC formulation to assess interference-induced performance degradation. Real-world and simulated results show that the jamming resistance $Q$ reliably predicts effective $C\!/\!N_0$ degradation across interference types.

\bibliography{ICL2026}
\bibliographystyle{IEEEtran}

\end{document}